\documentclass[prb,twocolumn,showpacs,preprintnumbers,amsmath,aps]{revtex4}
\usepackage{graphicx,hyperref}
\usepackage{bm}
\usepackage{subfigure}

\begin{document}

\title{GLASS TRANSITIONS MAY BE SIMILAR IN 2 AND 3 DIMENSIONS, AFTER ALL}

\author{Gilles Tarjus} \email{tarjus@lptmc.jussieu.fr}
\affiliation{LPTMC, CNRS-UMR 7600, Universit\'e Pierre et Marie Curie,
bo\^ite 121, 4 Pl. Jussieu, 75252 Paris c\'edex 05, France}

\date{\today}

\begin{abstract}
This is a commentary on two recent experimental papers in PNAS by Vivek et al. \onlinecite{vivek17} and Illing et al. \onlinecite{illing17} that convincingly address an issue at the junction of two fundamental questions in glass physics: the role of the dimensionality of space on the glass transition and the possible existence of long wavelength fluctuations in two-dimensional amorphous solids.
\end{abstract}

\maketitle

Understanding glasses and the glass transition is widely accepted as a deep, mysterious, and fundamental problem. Yet the consensus does not extend much further.  The topic is still hotly debated and progress toward a commonly accepted resolution seems slow for what is after all one of the oldest puzzles in physics. New theoretical tools and predictions do emerge, new phenomena are unveiled and clever experiments are nonetheless carried out. In this vein, the two recent experimental papers in PNAS by Vivek et al. \onlinecite{vivek17} and Illing et al. \onlinecite{illing17} convincingly address an issue at the junction of two fundamental questions in glass physics: the role of the dimensionality of space on the glass transition and the possible existence of long wavelength fluctuations in two-dimensional amorphous solids.

Is the nature of the glass transition different in 2 and 3 dimensions? Contrary to many ordering transition, such as crystallization, which are known to be different in two dimensions ($2D$) and in three dimensions ($3D$), there has been for some time a loose form of consensus that the glass transition is similar in $2D$ and $3D$. As summarized in a pithy sentence by P. Harrowell, ``in Flatland, glasses reproduce all the behaviour of their three-dimensional relatives'' \onlinecite{harrowell06}. The rationale behind this is that the glass transition involves no obvious long-range order nor spontaneous symmetry breaking. Actually, the experimentally observed glass transition is not even a true phase transition. It is a kinetic crossover, admittedly quite sharp, through which a liquid that, upon cooling, has become too viscous to flow and relax on a reasonable observation time (by anthropic standards) falls out of equilibrium. It then forms a glass, an ÒamorphousÓ solid whose structure looks as disordered as that of the liquid prior to the crossover \onlinecite{berthier11}.

However, a clear blow to this assumed similarity came from a recent comparative study of $2D$ and $3D$ model glass-forming liquids by computer simulation. Flenner and Szamel \onlinecite{flenner15} showed that the dynamics of $2D$ glass-formers is qualitatively different than that of their $3D$ counterparts. As illustrated in Fig. 1, the translational motion of the particles proceeds differently in 2 and $3D$. Particles stay trapped for relatively long times in the ``cage'' formed by their neighbors in $3D$, which gives rise to a plateau in the self-intermediate correlation function (Fig. 1, {\it Upper}). On the other hand, they can move sizable distances along with their neighbors with little change of their local structure in $2D$ (Fig. 1, {\it Lower}), which generates a strong dependence on the system size. Quite strikingly, in $2D$ but not in $3D$, as one cools the system, the translational motion, and the associated time-dependent correlation functions, appear to decouple from the motion of the particles involving a rearrangement of the local environment. The latter can be detected through ``bond-orientational'' correlation functions probing the orientational change of the vector between two nearest-neighbor particles. The authors then concluded that ``glassy dynamics in $2D$ and $3D$ are profoundly different''.

\begin{figure}[h]
\includegraphics[width=\linewidth]{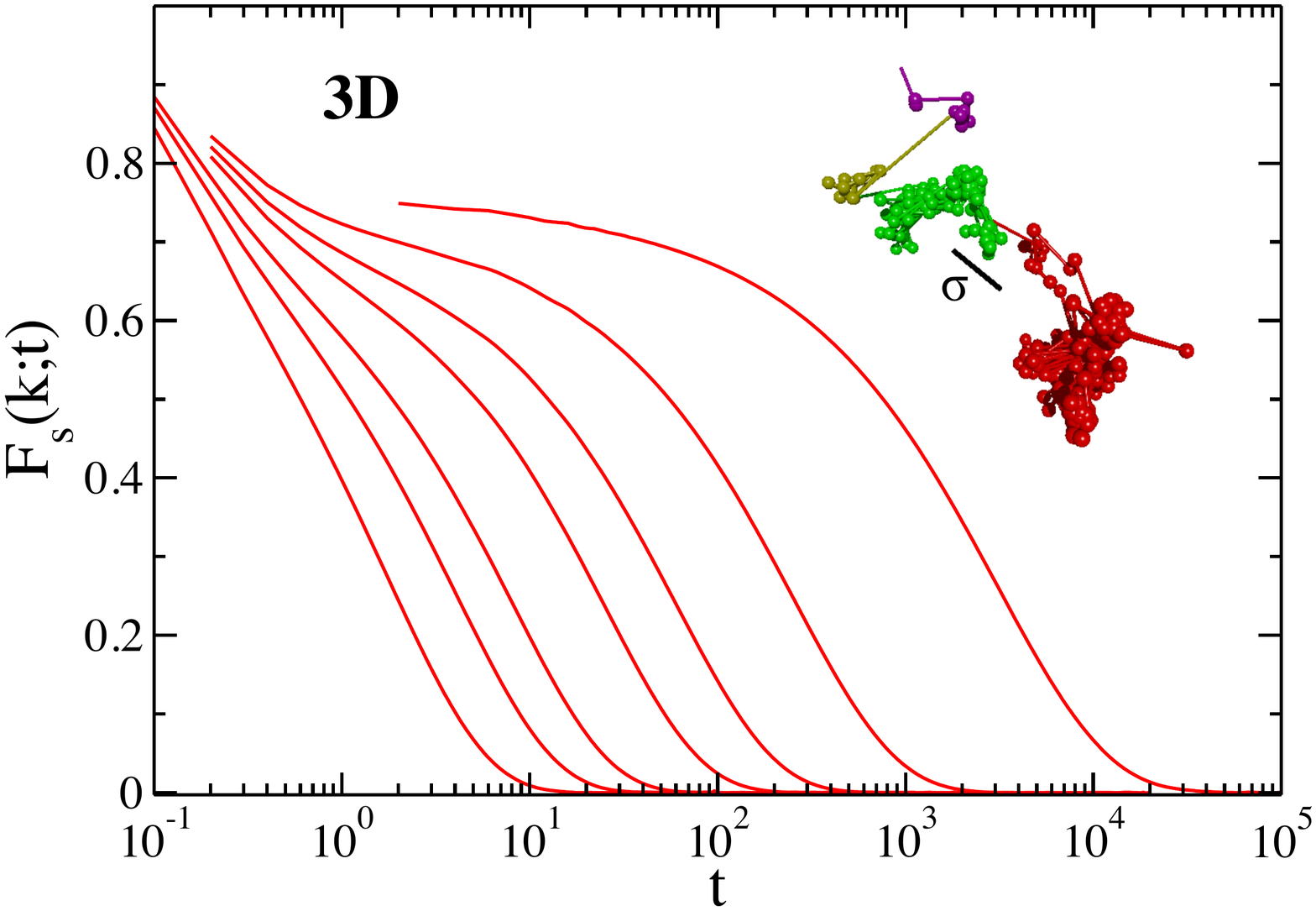}
\includegraphics[width=\linewidth]{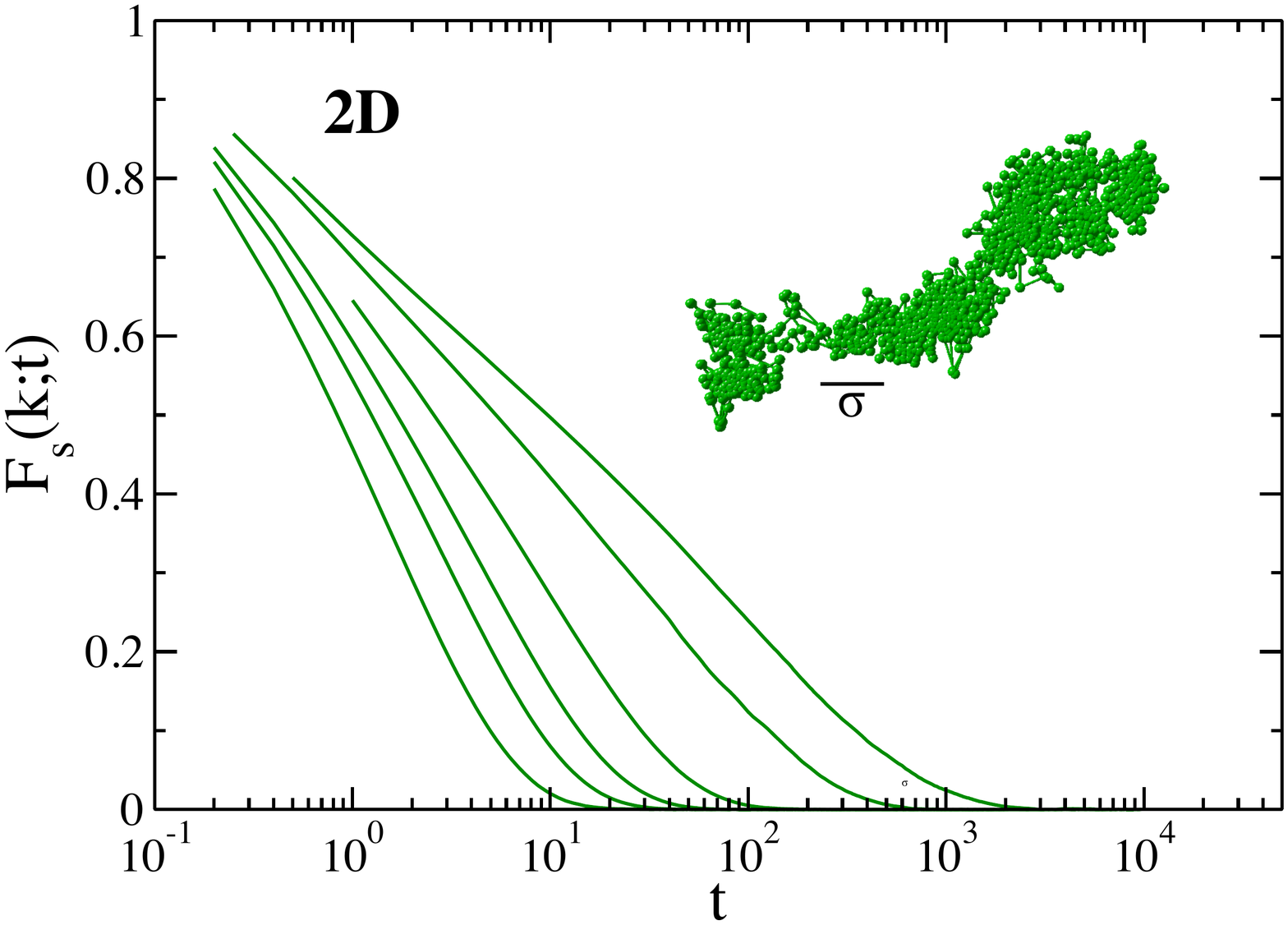}
\caption{Self correlation function of the density modes $F_s(k,t)$ versus time (log scale) in a $3D$ (top) and a $2D$ (bottom) glass-former for several temperatures $T$ (from left to right $T$ decreases). Inset: Trajectory plot of one particle at the lowest $T$. The results are from the computer simulation study of the Newtonian dynamics of model glass-forming liquids in Flenner and Szamel \onlinecite{flenner15}.}
\label{fig1}
\end{figure}

The two papers in PNAS \onlinecite{vivek17,Illing17} first present a beautiful experimental corroboration of the simulation results. To achieve this the two groups took advantage of the specific properties of soft (colloidal) matter: the big size of the colloids ($10^4$ to $10^6$ that of an atom) allows visualization through an optical microscope and the associated sluggish dynamics can be tracked and resolved in time, all of this at a particle level.

Settling the dimensionality dependence has both practical and fundamental benefits.  On the practical side, if no major change of the main physics takes place, reducing the dimensionality allows more convenient investigations and easier visualization of particle systems. At a fundamental level, space dimension can be used as an additional ``control'' parameter to gain insight into a specific physical phenomenon and disentangle the various mechanisms that may be at play. Experimentally, this can be undertaken in dimensions 3, 2, and sometimes 1. In a more abstract setting, theoretical physicists have come to consider (when possible) dimension as a parameter that can be continuously varied and taken to infinity. This is, for instance, a standard tool in the theory of phase transitions and critical phenomena. In the limit of an infinite number of dimensions, $D\to \infty$, a ``mean-field'' description, such as the Curie-Weiss theory of magnetism or the van der Waals theory of fluids, often becomes exact. Fundamentally, this stems from the suppression of spatial fluctuations in high dimensions, which allows a reduction of the problem at hand to that of a single constituent---for example, a particle or a spin---in the mean field created by all others. However, when decreasing the dimension $D$, the role of the spatial fluctuations increases and in low enough dimensions fluctuations become dominant and can even wipe out the phenomenon of interest (see also below). In many cases, the strong effect of fluctuations---and more to the point of long wavelength ones---is well accounted for by the renormalization group theory \onlinecite{wilson74}. 

A mean-field theory of the glass transition \onlinecite{kurchan12} has recently been established for fluids in the limit of an infinite number of dimensions. It is an unusually complex and elaborate construction that had been put forward before on less rigorous grounds \onlinecite{KTW89}. In this case fluctuations are also expected to become more important as dimension decreases and one reaches the physically relevant dimensions, $D=3$ and $D=2$. However, the precise nature of the relevant fluctuations remains elusive and their manifestation can a priori take a variety of forms \onlinecite{yaida16}. What is demonstrated in Vivek et al. \onlinecite{vivek17} and Illing et al. \onlinecite{illing17} is that a new type of fluctuations seems to emerge in $D=2$, which may explain most of the observed qualitative differences with glass formation in $3D$.

To discuss the nature of the spatial fluctuations seen in $2D$ but not in $3D$ glass-formers it is useful to first make a detour by $2D$ crystals. It was argued from heuristic arguments by Peierls \onlinecite{peierls34} and Landau \onlinecite{landau37} and then rigorously shown by Mermin \onlinecite{mermin66}, extending the earlier work of Mermin and Wagner \onlinecite{mermin68}, that there can be no long-range positional order at any nonzero temperature in $2D$ and less. Crystals in their conventional acceptation therefore do not exist in $2D$. The reason is that thermal fluctuations in the form of long wavelength density modes (acoustic phonons) lead to a divergent mean square displacement $<\Delta r^2>$ of the particles from their equilibrium positions, thereby destroying long-range periodic order. (Note that the argument applies to translational order but do not prevent long-range bond-orientational order.) In $2D$, this divergence is logarithmically slow with the system size $L$, $<\Delta r^2> \sim T \log(L/\sigma)$ where $\sigma$ is the inter-particle distance, which, as pointed out by Landau and Lifschitz, implies in practice that ``the size of the film for which the fluctuations are still small may be very great'' \onlinecite{landau80}. However, the fundamental importance of the result, namely that long wavelength fluctuations which one may refer to as Mermin-Wagner fluctuations destabilize the crystal [and other forms of long-range order as well \onlinecite{mermin68}] in $2D$, led Kosterlitz and Thouless to establish the existence of ``long-range topological order'' in such systems \onlinecite{KT73}, a far-reaching result that awarded them the Nobel Prize in Physics in 2016.

A glass, on the other hand, is a solid only for times shorter than that for relaxation to equilibrium and, to make matter worse, it is an amorphous solid that lacks periodicity. It has nonetheless been suggested that Mermin-Wagner-like density fluctuations could operate in $2D$ glass-formers.  There is indeed some evidence that at large scales beyond some characteristic length $\xi$, a glass behaves as a homogeneous elastic medium. The very same reasoning leading to the above divergence of the mean square displacement can be used with the inter-particle size $\sigma$ replaced by the possibly larger length $\xi$. This result however can only be valid for a limited duration since for elapsed times larger than the typical relaxation time the system should behave like a more conventional viscous liquid. Nonetheless, this suggests that vestigial Mermin-Wagner-like density fluctuations could  affect the dynamics of a $2D$ glass-former, more specifically the translational motion of the particles. 

A striking outcome of the two studies recently presented in PNAS \onlinecite{vivek17,illing17} is the evidence they both give that long wavelength density fluctuations are indeed present in $2D$ glass-formers and provide an additional channel for particle motion on top of the generic ``structural relaxation'' involving irreversible rearrangements of the local structure. This finding could explain why time-dependent translational correlation functions change on a faster time scale than correlation functions only sensitive to the rearrangements of the local environment of the particles (the ``cage'') such as bond-orientational ones. The two studies also offer a more direct test. From the knowledge of the individual particle trajectories the authors compute a variant of the translational correlation functions that is based on the displacement of the particles measured relative to that of the surrounding cage of neighbors. These new ``cage-relative'' functions are expected to be less sensitive to the long wavelength fluctuations that cause a displacement of the particles along with their environment. In $2D$ these functions indeed display a change in time that is now slower than the conventional translational functions \onlinecite{vivek17,illing17}, and Illing et al. \onlinecite{illing17} give indications that the associated time scale is comparable to that of the bond-orientational functions. In contrast,  the cage-relative and conventional translational correlation functions are not significantly different in $3D$. Carefully analyzed, Flatland ($2D$) studies could then, after all, provide insight into the generic features of glass formation. 

These experimental results are clearly important for glass physics. They are also stimulating as they raise new questions and open avenues for further experimental and numerical investigations. Among the fundamental issues triggered by the studies in Vivek et al. and Illing et al. \onlinecite{vivek17,illing17} are the theoretical foundations of these Mermin-Wagner fluctuations in $2D$ glass-formers and the possible interference with other types of fluctuations. As stressed above, the observed glass transition is not a bona fide phase transition and notions such rigidity and solidity for glasses still require some more robust theoretical underpinning. Whether the spectacular slowdown of dynamics upon cooling (or increasing the concentration) is a collective phenomenon controlled by some underlying but not observable phase transitions, where some form of long-range order settles in, is nonetheless a legitimate question. A quite different but complementary problem than that addressed in \onlinecite{vivek17,illing17} is the nature of the fluctuations that may destabilize these putative transitions and the ensuing ``lower critical dimension'' below which no such transition is possible as a result of too strong fluctuations. It is not at all clear that these fluctuations are long wavelength density modes as found in \onlinecite{vivek17,illing17}, nor that $D=2$ should then be the lower critical dimension. 

The strength of the soft-matter systems studied in Vivek et al. and Illing et al. \onlinecite{vivek17,illing17}, with big and slow colloidal particles that one can track in space and time at an individual level, comes with a down side for what concerns glass formation. The slowing down of dynamics with decreasing temperature or increasing concentration cannot be probed over more than $4$-$5$ orders of magnitude in relaxation time (much like computer simulation studies of liquid models). This should be contrasted with molecular glass-forming liquids for which a variation of the viscosity or the relaxation time of up to $14$ or $15$ orders of magnitude can be accessed. If glass formation involves the growth of spatial correlations of one sort or another beyond the mere interparticle distance, these correlations should be less pronounced in colloidal systems than in molecular liquids approaching their glass transition. How the various, possibly intertwined and/or competing, fluctuations affect the dynamics in the deeply supercooled---that is, highly viscous---regime of $2D$ glass-formers not accessible to colloidal systems then remains an open question.

\end{document}